\documentclass[aip,twocolumn,letterpaper]{revtex4}

\usepackage{fullpage}

\usepackage{graphics}
\usepackage{epsfig}

\usepackage{xcolor}

\usepackage[margin=1.0in]{geometry}

\begin{document}
\title{Local analysis of fast magnetic reconnection}
\author{Allen H Boozer}
\affiliation{Columbia University, New York, NY  10027\\ ahb17@columbia.edu}

\begin{abstract}

Fast magnetic reconnection is defined by the topology of the magnetic field lines changing on a timescale that is approximately an order of magnitude longer than the topology-conserving ideal-evolution timescale.  Fast reconnection is an intrinsic property of Faraday's law when the evolving magnetic field depends non-trivially on all three spatial coordinates and is commonly observed---even when the effects that allow topology breaking are arbitrarily small.  The associated current density need only be enhanced by a factor of approximately ten and flows in thin but broad ribbons along the magnetic field.  These results follow from the variation in the separation of neighboring pairs of magnetic field lines, which in an ideal evolution typically increases exponentially with time, and the existence of a spatial scale below which magnetic field lines freely change their identities due to non-ideal effects such as resistivity.  Traditional reconnection theory ignores exponentially large variations and relies on the current density reaching a magnitude that is exponentially larger than is actually required.  Here, an analysis of the behavior of magnetic field lines in the neighborhood of an arbitrarily chosen line is used to obtain more precise and rigorous results on intrinsic reconnection.  The maximum parallel kinetic energy of collisionless charged particles is shown to have an exponential increase in time during a generic magnetic evolution.

\end{abstract}

\date{\today} 
\maketitle


\section{Introduction}

The evolution equation for magnetic fields, using the simple Ohm's law, $\vec{E}+\vec{v}\times\vec{B}=\eta\vec{j}$, is
\begin{equation}
\frac{\partial \vec{B}}{\partial t}= \vec{\nabla}\times\left(\vec{v}\times\vec{B} -\frac{\eta}{\mu_0}\vec{\nabla}\times\vec{B}\right).
\end{equation}
This equation is of the advection-diffusion form as is the equation for the equilibration of the temperature in a room \cite{Boozer:rec-phys},
\begin{equation}
\frac{\partial T}{\partial t}= -\vec{\nabla}\cdot\left(\vec{v}T - D_T\vec{\nabla}T\right).
\end{equation}
In standard problems of spatial scale $a$, the advective is enormous compared to the diffusive term.  This ratio is given by the magnetic Reynolds number $R_m\equiv va/(\eta/\mu_0)$ or the P\'eclet number $P_e\equiv va/D_T$, which are often between $10^4$ and $10^{10}$.  Empirically, magnetic reconnection and thermal equilibration are observed to occur on a timescale approximately an order of magnitude longer than the ideal evolution time $a/v$ regardless of how small the diffusive term may be.   Once understood, the obvious explanation is that neighboring streamlines of natural flows, $d\vec{x}/dt=\vec{v}$, generally have a spatial separation that increases exponentially with time, which makes the advection-diffusion equation exponentially sensitive to the diffusive terms.   This paradigm has been accepted for thermal equilibration since the 1984 paper by Aref \cite{Aref:1984} but has been largely ignored without justification by the magnetic-reconnection community.


 Faraday's law, $\partial\vec{B}/\partial t=-\vec{\nabla}\times\vec{E}$, determines all evolutions of magnetic fields---even the most paradoxical, a fast magnetic reconnection, a topology change on a timescale that is approximately an order of magnitude longer than the topology-conserving ideal timescale. 
 An understanding of fast magnetic reconnection requires an understanding of the  mathematical properties of Faraday's law and the electric field, but many of the most important properties are little known.  Eight of these were reviewed in Appendix A of Reference \cite{Boozer:surf2022}.

The first property in this list of eight is a general expression for the electric field.  Conventionally, the electric field in Faraday's law is expressed in a frame of reference moving with the velocity $\vec{v}$, the mass-flow velocity of the plasma.  This is called Ohm's law.  In its simplest form, $\vec{E}+\vec{v}\times\vec{B}=\eta\vec{j}$, but realistic expressions are far more complicated.  A different procedure was introduced by Boozer \cite{Boozer:coordinates} in 1981.  The velocity of the frame of reference $\vec{u}_\bot$ in the non-relativistic Lorentz transformation of electromagnetic fields is chosen to maximally simplify the general expression for $\vec{E}$.  One obtains
\begin{eqnarray}
\vec{E}+\vec{u}_\bot\times\vec{B}= - \vec{\nabla}\Phi +\mathcal{E}\vec{\nabla}\ell. \label{E-rep}
\end{eqnarray}
The distance along a magnetic field line is $\ell$.  The field-line constant $\mathcal{E}$ is required to make $\Phi$ single valued; any part of $\mathcal{E}$ that is not independent of $\ell$ can be absorbed into $\Phi$.  In a torus, $\mathcal{E}$ is closely related to the loop voltage.  In systems in which field lines strike boundaries, $\mathcal{E}$ is required to ensure $\int\vec{E}\cdot d\vec{\ell}$ obeys the boundary conditions; $d\vec{\ell}\equiv \hat{b} d\ell$ with $\hat{b}\equiv\vec{B}/B$.  

Where $\vec{B}\neq0$, the validity of Equation (\ref{E-rep}) is essentially obvious.  The part of $\vec{E}$ that is parallel to $\vec{B}$ can be represented by the parallel component of $- \vec{\nabla}\Phi +\mathcal{E}\vec{\nabla}\ell$, and the part of $\vec{E}$ that is perpendicular to $\vec{B}$ can be represented by the choice of $\vec{u}_\bot\times\vec{B}$.  The analysis can be extended to magnetic fields that have nulls at discrete points.   A small sphere is placed around each place where $\vec{B}=0$ with the potential on each sphere chosen so no net current enters the sphere, $\oint\vec{j}\cdot d\vec{a}=0$. This defines a boundary condition on the integral $\int \vec{E}\cdot d\vec{\ell}$.  This extension is discussed in Section IV of Reference \cite{X-null} and in Elder and Boozer \cite{Elder:2021}.  Only point nulls are of practical interest since a line null can be removed or turned into well separated point nulls by an arbitrarily small perturbation.

Equation (\ref{E-rep}) provides a clear differentiation between an ideal, $\mathcal{E}=0$,  and a non-ideal, $\mathcal{E}\neq0$, magnetic evolution.   When $\mathcal{E}=0$,  $\vec{u}_\bot$ is the velocity of the magnetic field lines.  In 1958 Newcomb \cite{Newcomb} explained the distinction between $\vec{u}_\bot$ and the plasma velocity $\vec{v}$.  This is the second property in the list of eight reviewed in Appendix A of Reference \cite{Boozer:surf2022}.  Newcomb showed the magnetic flux enclosed by a closed contour that moves with the velocity $\vec{u}_\bot$ is conserved, but this conservation law need not hold when the contour moves with the  plasma velocity,  $\vec{v}$.  When $\mathcal{E}\neq0$, the evolution of magnetic field lines is no longer given by $\vec{u}_\bot$, Reference  \cite{Boozer:rec-phys}.

Magnetic reconnection was defined in 1956 by Parker and Krook \cite{Parker-Krook:1956} as the ``\emph{severing and  reconnection of lines of force.}"  In other words, magnetic reconnection is a topology change in the field lines, which, as proven by Newcomb, requires a non-ideal evolution.  When the electric field is written in the form of Equation (\ref{E-rep}), a topology change requires $\mathcal{E}\neq0$.

The ratio of the non-ideal to the ideal-evolution drive for a magnetic evolution is
\begin{equation}
\epsilon_{ni}\equiv\frac{\big< \big|\mathcal{E}\big| \big>}{\big<\big| \vec{\tilde{u}}_\bot\times\vec{B} \big|\big>},
\end{equation}
where $<\cdots>$ is a spatial average and the variation in the velocity is $\vec{\tilde{u}}\equiv \vec{u}_\bot -\big<\vec{u}_\bot\big>$.  

The ubiquity of fast magnetic reconnection raises two questions.  (1) How can a timescale set by the ideal evolution velocity $\vec{u}_\bot$ be the primary determinant of the onset and the speed of reconnection when $\epsilon_{ni}$ is arbitrarily small?  (2) How can a process that leads to a fast magnetic reconnection be sufficiently intrinsic to explain its ubiquity?

The focus of this paper is more rigorous derivations of the intrinsic properties of Faraday's law.  

What can and will be addressed with remarkable completeness is the behavior of neighboring magnetic field lines, which means lines that are spatially separated by a distance small compared to the spatial scale $a_u$  of the velocity of the magnetic field lines, $\vec{u}_\bot$.  Numerical studies \cite{Rec-example} show that the properties of neighboring lines provide results that qualitatively describe reconnection effects out to the scale $a_u$.  Indeed, one expects the natural scale of the region in which reconnection takes place to extend a distance that approximately equals $a_u$ across the magnetic field lines.  

Section \ref{sec:fast rec}, \emph{Fast magnetic reconnection}, reviews the various theories of magnetic reconnection.

Section \ref{sec:near given}, \emph{The magnetic field near a given field line}, derives the equations for the trajectories of magnetic field lines in the neighborhood of an arbitrarily chosen line.  These equations were derived in Reference \cite{Boozer:B-line.sep} in 2012, but Section \ref{sec:near given} places these equations in the form that is required for studying an ideal evolution.

Section \ref{sec:ideal evolution}, \emph{Ideal evolution of neighboring lines}, extends the results of Section \ref{sec:near given} to an arbitrary ideal magnetic-evolution.  The results cannot be directly extended to non-ideal evolutions for then the central field line does not remain intact.  Nevertheless, the most important feature for fast magnetic reconnection can be studied---the evolution of $\Delta_{max}/\Delta_{min}$, the ratio of the maximum to minimum separation between neighboring magnetic field lines.

Section \ref{sec:footpoint},  \emph{Reconnection due to footpoint motion}, discusses the implications of Section \ref{sec:ideal evolution} for reconnection driven by the motion of the interception points of magnetic field lines with a flowing perfectly-conducting boundary.

Section \ref{sec:surface distortion},  \emph{Ideal evolution of magnetic surfaces}, derives an equation that shows the evolution of the maximum to the minimum separation between magnetic surfaces has a natural exponential dependence on time. 

Section \ref{sec:discussion}, \emph{Discussion} gives an overview of the paper and the most important new results. 

Appendix \ref{sec:acceleration}, \emph{Exponential particle acceleration}, shows that when a magnetic field is evolving---even evolving ideally---the maximum parallel kinetic energy of collisionless charged particles generally increases exponentially with time.  

Appendix \ref{sec:canonical coord}, \emph{Magnetic canonical coordinates},  gives a general canonical representation of magnetic fields that is not restricted to the neighborhood of a particular magnetic field line but is closely related to the representation of Section \ref{sec:near given} that is.

Appendix  \ref{sec:Spitzer}, \emph{Spitzer's stellarator fields}, discusses the curl-free magnetic fields given by Spitzer in his foundational paper on stellarators.  In particular, it is explained why the torsion of a magnetic field line is second order in the distortion from a straight field line and that the quadrupole field discussed in Section \ref{sec:near given} can produce a net twist of the magnetic field lines in second order.


\section{Fast magnetic reconnection \label{sec:fast rec}}

\subsection{Intrinsic fast magnetic reconnection \label{sec:intrinsic} }


In the limit of non-ideality becoming arbitrarily small, $\epsilon_{ni}\rightarrow0$, a pair of magnetic field lines must pass within a very small distance of each other, $\Delta_d\propto\epsilon_{ni}$, to lose their distinguishability and reconnect.  

Nevertheless, when a pair of field lines have become indistinguishable anywhere along their trajectories, they have also become indistinguishable everywhere along their trajectories---no matter how great their separation may become.  In other words, magnetic reconnection occurs over a large spatial scale when the separation between a pair of magnetic field lines undergoes sufficiently large changes along the length of the lines,  $\Delta_{max}/\Delta_{min}$, the ratio of the maximum to the minimum separation. 

Three distance scales characterize reconnection: The first is the distance scale $\Delta_d$ at which field lines become indistinguishable due to direct non-ideal effects.  The second is the distance scale $a_r$ over which reconnection must occur to have a significant effect on the evolution and to be of practical importance. The third is the distance scale $a_u$ of spatial variations in $\vec{u}_\bot$.

Plasma resistivity causes magnetic field lines to interdiffuse, become indistinguishable, and reconnect over the scale $\Delta_d=\sqrt{(\eta/\mu_0)\tau_\eta}$ during the time $\tau_\eta$.  The shortest timescale $\tau_\eta$ defines the shortest distance over which lines can remain distinguishable, and that timescale is $\tau_\eta=\Delta_d/\tilde{u}_\bot$.  Consequently, resistivity alone gives $\Delta_d\sim  \eta/(\mu_0\tilde{u}_\bot)$.  Other non-ideal effects, including the grid scale in numerical simulations, contribute to determining the minimum scale $\Delta_d$ over which pairs of magnetic field lines can be distinguished.  

Naively, one would expect reconnection to be irrelevant when the distance required for the distinguishability of the lines, $\Delta_d$, is tiny compared to the spatial scale for reconnection to be significant, $a_r$, but this is clearly false if the ratio of separations between pairs of magnetic field lines, $\Delta_{max}/\Delta_{min}$, becomes as large as $a_r/\Delta_d$ with $a_r/\Delta_d\sim1/\epsilon_{ni}$.

The spatial scale $a_u$ of variations in $\vec{u}_\bot$ is important because the maximum separation $\Delta_{max}$ can increase exponentially in time for field lines with a minimum separation $\Delta_{min}=\Delta_d$ only when $\Delta_{max}\lesssim a_u$.  When the scale of a chaotic flow $a_u$ is large compared to $\Delta_{max}$, an initially circular tube of magnetic flux distorts into an ellipse with a large diameter $\Delta_{max}$ that increases exponentially with time and a small diameter $\Delta_{min}$ that decreases exponentially with time.  However, when $\Delta_{max}$ becomes large compared to the scale of the chaotic flow $a_u$, the evermore elongated tubes of magnetic flux develop folds, and the maximum separation increases only diffusively, which in simple cases means as $\sqrt{t}$. 

The natural scale for fast magnetic reconnection across the magnetic field lines is $a_u$.  The importance of fast magnetic reconnection is determined by the relative size of $a_u$ and $a_r$.  Natural flows may have many spatial scales, and the scales that are comparable to or longer than $a_r$ are of greatest practical importance.  Extremely short scales, as in turbulent flows, are only important for rapid reconnection when the flow speed of the turbulent eddies is far larger than the speed of the longer scale parts of the flow \cite{Boozer:rec-phys}.

The 1984 paper of Aref \cite{Aref:1984} showed that neighboring streamlines of flows, such as $\vec{u}_\bot$, intrinsically separate exponentially in time, which implies the natural behavior of $\ln(\Delta_{max}/\Delta_{min})\sim t/\tau_{ev}$, where $\tau_{ev}$ is the timescale of the ideal evolution.  In other words, the natural timescale for the onset of a large scale magnetic reconnection is $\sim\tau_{ev}\ln(a_r/\Delta_d)$.  In systems of interest, $\ln(a_r/\Delta_d)\sim\ln(1/\epsilon_{ni})$ is generally of order ten, but certainly less than a hundred.

Aref's work explained the ubiquity of temperature relaxation in a room in tens of minutes rather than the few weeks that would be required by diffusive transport.  The flow that causes this relaxation is constrained by an energy argument to be divergence free, which requires the region of relaxation have at least two non-trivial spatial dimensions \cite{Boozer:rec-phys}.  For magnetic reconnection, an analogous energy argument requires the region of relaxation have at least three non-trivial spatial dimensions \cite{Boozer:rec-phys}.  This essentially follows from the requirement that the flow involve at least two directions with the magnetic field pointing in a third.

The current density required to achieve $\Delta_{max}/\Delta_{min}\sim a_r/\Delta_d$ scales \cite{Boozer:B-line.sep} as  $\ln(a_r/\Delta_d)$ and lies in numerous thin but broad ribbons throughout the region of magnetic reconnection \cite{Boozer:rec-phys}.  This was illustrated for a smooth large-scale flow of the field lines $\vec{u}_\bot$ by Boozer and Elder  \cite{Rec-example}.

\subsection{Traditional reconnection theory}

The traditional theory of magnetic reconnection, which developed out of two-dimensional models, ignores exponentially large variations in separation between the magnetic field lines in neighboring pairs.  In two dimensions, an exponential increase in $\Delta_{max}/\Delta_{min}$ for pairs of field lines throughout a volume is impossible \cite{Boozer:rec-phys}.  The maximum current density required in traditional theory scales as $1/\epsilon_{ni}$ rather than as the $\ln(1/\epsilon_{ni})$ scaling that arises when exponentiation is taken into account.   A feeling for the ratio of $1/\epsilon_{ni}$ to $\ln(1/\epsilon_{ni})$  in the solar corona is gained by noting that this ratio is comparable to the ratio of the age of the universe to the life expectancy of a person.

Traditional reconnection theory explains fast magnetic reconnection by assuming that, even though the non-ideality, $\epsilon_{ni}$ is extremely small, the actual reconnection takes place in a thin layer in which
\begin{eqnarray}
 \frac{\big|\mathcal{E} \big|}{\big<\big| \vec{\tilde{u}}_\bot\times\vec{B} \big|\big>}\approx1.  
\end{eqnarray}
This was the solution given in 1988 by Schindler, Hesse, and Birn \cite{Schindler:1988}.  In their model, $\mathcal{E}$ is given by a field-line average of $\eta j_{||}$. The thickness of the layer in which the current must be concentrated, $\Delta_\eta$, is related to the reconnection scale $a_r$ by
\begin{eqnarray}
\frac{a_r}{\Delta_\eta} \approx \frac{\mu_0\tilde{u}_\bot a_r}{\eta} \equiv R_m.
\end{eqnarray}
$R_m$ is called the magnetic Reynolds number and can be greater than $10^{10}$ in coronal problems. The current density in the thin layer is approximately $R_m$ times larger than the characteristic current density, which is given by the scale $a_r$.  

\begin{figure}
\centerline{ \includegraphics[width=2.5in]{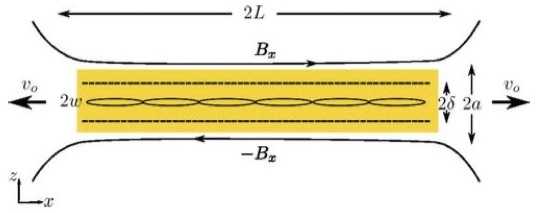}}
\caption{In plasmoid models, oppositely directed fields in the $B_x$ direction are pushed together forming a narrow current sheet.  Tearing instability of the sheet current creates the plasmoids, which are expelled at the Alfv\'en speed from the two ends of the current sheet. Reproduced from Y.-M. Huang, L. Comisso, and A. Bhattacharjee, Physics of Plasmas \textbf{26}, 092112 (2019), with the permission of AIP Publishing.}
\label{fig:plasmoid}
\end{figure}

Traditional reconnection theory \cite{Hesse-Cassak2020}, including plasmoid theory \cite{plasmoid}, Figure \ref{fig:plasmoid}, has not focused on how such an intense current can be produced in a timely manner, but rather how it can be maintained assuming its presence as an initial condition.  

As discussed in Section \ref{sec:surface distortion}, one can find cases in which the current density will become singular on a surface in the limit as time goes to infinity, but the required time can be too long to explain fast reconnection.  The required time for a current to concentrate in a thin layer about a surface depends on how long is required for a shear Alfv\'en wave propagating along magnetic field lines to cover the region enclosed by this layer \cite{Rec-example,Boozer:surf2022}.  

What is the range of applicability of traditional reconnection theory?  In systems with only two non-trivial spatial coordinates, there is no real alternative.  But, the applicability of traditional reconnection theory to three-dimensional systems  is far from obvious when the magnetic Reynolds number $R_m$ is large.  The intrinsic reconnection described in Section \ref{sec:intrinsic} would be expected to arise and fundamentally change the magnetic evolution when the current density is smaller by a factor $R_m/\ln{R_m}\approx 4\times10^8$ when $R_m=10^{10}$.

  
 Understanding fast reconnection as an intrinsic implication of Faraday's law is a sharp shift in paradigm from traditional reconnection theory.  The concept of a paradigm shift in science was introduced by Thomas Kuhn \cite{Kuhn} in 1962, \emph{one of the most influential works of history and philosophy written in the 20th century} \cite{Britannica:Kuhn}.
 
 The shift in paradigm between traditional reconnection theory and the intrinsic implications of Faraday's law produces changes (1) in questions, (2) in concepts, and (3) in standards of validation.  These are the standard changes of any paradigm shift.
 \vspace{0.1in}
 
 \textbf{1.  Changes in questions}
  \vspace{0.1in}
 
 a. How can an arbitrary magnetic field naturally and sufficiently rapidly evolve into a state of fast magnetic reconnection?  This question is neither a focus nor convincingly addressed in traditional reconnection theory.
  \vspace{0.1in}
 
 b. What are the implications of different properties along the direction perpendicular to the page in Figure \ref{fig:plasmoid}, the $y$ direction?  The $y$ direction could have perfect symmetry from plus to minus infinity or have perfect symmetry in a periodic angle.  The $y$ direction could also be bounded by a perfectly conducting or by an insulating wall.  This question is neither a focus nor convincingly addressed in traditional reconnection theory.
   \vspace{0.1in}
   
  c.  Over what regions can particles be accelerated?  Appendix \ref{sec:acceleration} shows that when a magnetic field is evolving---even evolving ideally---the maximum parallel kinetic energy of collisionless charged particles naturally increases exponentially with time.   
  
  This follows from Hamiltonians of the $H(p,q,t)$ form generically having chaotic trajectories, which means neighboring pairs of trajectories $p(t),q(t)$ separate exponentially in time.  The simplest Hamiltonian for charged particle motion along a magnetic field line is $H_p(p_{||},\ell,t) = (p_{||}^2/2m) +\mu B(\ell,t)$ is of the $H(p,q,t)$ form when the magnetic field is evolving.

 The acceleration region is spread through a large volume in intrinsic reconnection unlike in the forms of traditional reconnection theory in which the primary acceleration mechanism is taken to be the non-ideal electric field $\mathcal{E}\vec{\nabla}\ell$, which is extremely large but only in thin sheets. 

Although the dominant acceleration mechanism was traditionally thought to be from the non-ideal part of the electric field, Dahlin, Drake, and Swisdak noted in 2016 that the parallel electric field is not an efficient mechanism for particle acceleration.   In the next year, they discussed an alternative mechanism  \cite{Drake:2017}.  Boozer gave the analogous theory of particle acceleration  in 2019 for intrinsic reconnection  \cite{Boozer:acc}.  The derivations given in these three papers are subtle and complicated unlike the demonstration given in Appendix \ref{sec:acceleration}.
    \vspace{0.1in}
    
\textbf{2. Changes in concepts}
\vspace{0.1in}

a.  The primary conceptual failure in traditional reconnection theory is the neglect of the dominant effect in making Faraday's law exponentially sensitive to non-ideal effects.  This is the characteristic exponential increase in the ratio between the maximum and minimum separation of neighboring pairs of magnetic field lines.
\vspace{0.01in}

b.   Because the dominant effect is neglected, traditional reconnection theory has focused on the maintenance of the extremely large electric field thought to be required by Maxwell's equations.  In three-dimensional reconnection, the electric field need not be large and the region in which reconnection actually takes place need not be thin---a fundamental conceptual change. 
\vspace{0.1in}

c.  Traditional reconnection theory fails to distinguish the plasma and the magnetic-field-line velocities.  This distinction, which has been known since the 1958 paper of Newcomb \cite{Newcomb}, greatly simplifies Faraday's law, and clarifies the relative importance of different plasma effects to reconnection.
\vspace{0.1in}

\textbf{3.  Changes in standards of validation}
\vspace{0.1in}

a.  A valid theory of magnetic reconnection should be implied by mathematics coupled with Maxwell's equations.  Despite the traditional belief, there is no implication that a large electric field is required to produce a fast reconnection when there is a non-trivial dependence on all three spatial coordinates.  What is implied by mathematics and Maxwell's equations is that the magnetic-field-line velocity of an ideal evolution will generally make a magnetic evolution exponentially sensitive to non-ideal effects \cite{Boozer:rec-phys}.
\vspace{0.001in}
	
b.  The observation of current sheets is often assumed to validate traditional reconnection theory.  However, the existence of thin but wide ribbons of current is characteristic of a near ideal evolution in a system with three spatial coordinates \cite{Boozer:rec-phys,Rec-example}.  This is true even when the  $\vec{u}_\bot$ that defines the ideal evolution is a simple non-turbulent flow \cite{Rec-example}.  The maximum current density along the magnetic field, $j_{||}$, is exponentially larger in traditional theory in comparison to what is expected from Maxwell's equations in a generic evolution.  An important observational question is whether there are actual measurements of the extreme values of $j_{||}$ that traditional theory predicts.
\vspace{0.01in}
	
c.  As discussed in Appendix \ref{sec:acceleration}, particle acceleration can occur even when the non-ideal part of the electric field $\mathcal{E}$ is zero.  Consequently, particle acceleration does not demonstrate the existence of a large $\mathcal{E}$.  Studies of intrinsic reconnection are needed in order to determine its observational implications on particle acceleration.
\vspace{0.1in}

As discussed by Thomas Kuhn, paradigm changes are always met with great resistance.  The practitioners of the old paradigm naturally cling to a method of thought that has long been accepted as a basis of truth upon which careers, publications, and grants can be based.  Nevertheless, truth in science is not demonstrated by history or a majority vote of scientists.



\subsection{Other reconnection theories}

Two other theories of magnetic reconnection should be discussed:  (1) plasma turbulence and (2) theories associated with Eric Priest.

\subsubsection{Turbulence models}
 
 Turbulence can enhance the rate of reconnection.  As discussed in Section VIII of Reference \cite{Rec-example}, turbulence is of two types, (1) an enhanced effective resistivity to $j_{||}$, which would make $\Delta_d$ larger and  (2) a turbulent velocity of magnetic field lines.  
 
Micro-turbulence, which is turbulence on a spatial scale described by the Fokker-Planck equation for plasmas, could produce a greatly enhanced effective resistivity to $j_{||}$.  However, micro-turbulence has had limited discussion in the reconnection literature.
 
There is an extensive literature on the enhancement of reconnection by turbulence in the magnetic field line velocity \cite{Lazarian:1999,Eyink:2011,Eyink:2015,Matthaeus:2015,Matthaeus:2020}, which is reviewed in \cite{Lazarian:2020rev}.  Turbulence can not directly break magnetic field lines.  Turbulence can slow the increase in $\Delta_{max}$ in the $\Delta_{max}/\Delta_{min}$ ratio from an exponential to a square-root dependence on time when the spatial scale of the turbulent eddies is small compared to the required scale $\Delta_{max}\approx a_r$ for reconnection to be important \cite{Boozer:rec-phys}.


\subsubsection{Models by Eric Priest}

Eric Priest has inspired a large body of work on three-dimensional structures that tend to concentrate currents and thereby lead to enhanced reconnection \cite{Priest:2016}.  These structures are null points, where the magnetic field vanishes, separators, which are magnetic field lines joining null points, and quasi-separatrix layers, which are regions in a magnetic field where the gradient of a footpoint mapping is large. Quasi-separatrix layers are related to papers by Boozer and Elder \cite{Rec-example} and by Reid, Parnell, Hood, and Browning \cite{Reid:2020}, which discuss two closely related models of reconnection in the solar corona.



\section{The magnetic field near a given field line \label{sec:near given} }

Magnetic field lines are defined at fixed points in time.  In an ideally evolving magnetic field,  the ratio of the maximum to the minimum separation, $\Delta_{max}/\Delta_{min}$, between neighboring magnetic field lines can become exponentially large through a finite volume as time evolves, which gives a fast magnetic reconnection when the exponentiation is sufficient to counterbalance the smallness of the non-ideal effects that directly make field lines that come closer than $\Delta_d$ indistinguishable.

The implication is that important reconnection properties can be studied by examining the behavior of magnetic field lines that are separated from an arbitrarily chosen line $\vec{x}_0(s)$ by a distance $\rho$ as $\rho\rightarrow0$.  A Hamiltonian description of these neighboring field lines was first given in  Reference \cite{Boozer:B-line.sep}, and the relation between the Hamiltonian and the current density $j_{||}$ flowing along the line $\vec{x}_0$ was also found.  These equations will be rederived in this section in a convenient form for studying field-line evolution.  The arbitrarily chosen magnetic field line $\vec{x}_0(s)$ is the central field line of the analysis.

Section \ref{sec:ideal evolution} extends the results on nearby, $\rho\rightarrow0$, lines to ideally evolving magnetic fields.  The extension is needed to study the evolution in the separation of neighboring magnetic field lines and of the current density $j_{||}$ along the central line.


\subsection{Coordinates near a magnetic field line \label{sec:intrinsic coord} }

Positions near an arbitrarily chosen magnetic field line, $\vec{x}_0(s)$, which means $d\vec{x}_0/ds\equiv\hat{b}(\vec{x}_0)$, can be described using the intrinsic coordinates of Courant and Snyder \cite{Courant:1958};
\begin{equation}
\vec{x}_I(\rho,\alpha,s)=\rho\cos\alpha \hat{\kappa}_0 + \rho \sin\alpha \hat{\tau}_0 + \vec{x}_0(s). \label{Intrinsic coordinates}
\end{equation}
The unit vector $\hat{b}$ is defined as  $\hat{b}\equiv\vec{B}/B$, and $s$ has the interpretation as the distance along the line $\vec{x}_0(s)$.

A smooth curve $\vec{x}_0(s)$, which is parameterized by the distance along the curve $s$, can be used to define three orthonormal unit vectors: the tangent unit vector $\hat{b}_0(s)\equiv\hat{b}(\vec{x}_0)$, the curvature unit vector $\hat{\kappa}_0$, and the torsion unit vector $\hat{\tau}_0$.  The definitions are
\begin{eqnarray}
\frac{d\vec{x}_0}{ds} &=&\hat{b}_0; \\
\frac{d\hat{b}_0}{ds}&=&\kappa_0\hat{\kappa}_0; \label{curv def}\\
\frac{d\hat{\kappa}_0}{ds}&=&-(\kappa_0 \hat{b}_0+\tau_0\hat{\tau}_0); \label{torsion def}\\
\frac{d\hat{\tau}_0}{ds}&=&\tau_0\hat{\kappa}_0;\\
\hat{b}_0 &=& \hat{\kappa}_0\times\hat{\tau}_0. \label{vec-prod}
\end{eqnarray} 
The signs of the curvature $\kappa_0$ and the torsion $\tau_0$ are chosen so Equation (\ref{vec-prod}) has the standard sign.  These relations, which were first derived by Jean Fr\'ed\'eric Frenet in 1847 and independently by Joseph Alfred Serret in 1851, are easily obtained.  

By their definition, orthogonal unit vectors satisfy $\hat{e}_i\cdot\hat{e}_j=\delta_{ij}$, so  $(d\hat{e}_i/ds)\cdot\hat{e}_j+\hat{e}_i\cdot (d\hat{e}_j/ds)=0$.  Differentiating $\hat{b}_0\cdot\hat{b}_0=1$, one finds $d\hat{b}_0/ds$ must have the form $d\hat{b}_0/ds=\kappa_0\hat{\kappa}_0$, where $\hat{\kappa}_0$ must be orthogonal to $\hat{b}_0$.  Differentiating $\hat{\kappa}_0\cdot\hat{b}_0=0$, one finds the form $d\hat{\kappa}_0/ds=-(\kappa_0 \hat{b}_0+\tau_0\hat{\tau}_0)$.  Finally differentiating $\hat{\tau}_0\cdot\hat{b}_0=0$ and $\hat{\kappa}_0\cdot\hat{\tau}_0=0$, one finds that $d\hat{\tau}_0/ds=\tau_0\hat{\kappa}_0$.

The derivatives of the position vector $\vec{x}_I(\rho,\alpha,s)$ with respect to the three coordinates are called the tangent vectors.  Using the relations obeyed by the Frenet unit vectors:
 \begin{eqnarray}
\frac{\partial\vec{x}_I}{\partial \rho}&=& \cos\alpha\hat{\kappa}_0+\sin\alpha\hat{\tau}_0\equiv\hat{\rho};\\
\frac{\partial\vec{x}_I}{\partial \alpha}&=& \rho\left(\cos\alpha\hat{\tau}_0-\sin\alpha\hat{\kappa}_0\right)\equiv\rho\hat{\alpha};\\
\frac{\partial\vec{x}_I}{\partial s}&=& (1-\rho\kappa_0\cos\alpha)\hat{b}_0-\rho\tau_0\hat{\alpha}.
\end{eqnarray}
The coordinate Jacobian is
\begin{eqnarray}
\mathcal{J}_I&\equiv& \left(\frac{\partial\vec{x}_I}{\partial \rho}\times\frac{\partial\vec{x}_I}{\partial \alpha}\right)\cdot\frac{\partial\vec{x}_I}{\partial s} \\
&=& \frac{1}{(\vec{\nabla}\rho\times \vec{\nabla}\alpha)\cdot \vec{\nabla}s}\\
&=&(1-\rho\kappa_0\cos\alpha)\rho.
\end{eqnarray}

The appendix to Reference \cite{Boozer:RMP} gives a two-page derivation of the theory of general coordinates.  This includes the dual relations,
\begin{eqnarray}
\vec{\nabla}\rho&=&\frac{1}{\mathcal{J}_I}\frac{\partial\vec{x}_I}{\partial \alpha}\times\frac{\partial\vec{x}_I}{\partial s}=\hat{\rho}; \\
\vec{\nabla}\alpha&=&\frac{1}{\mathcal{J}_I}\frac{\partial\vec{x}_I}{\partial s}\times\frac{\partial\vec{x}_I}{\partial \rho}=\frac{\hat{\alpha}}{\rho}+\frac{\tau_0\hat{b}_0}{1-\rho\kappa_0\cos\alpha}; \hspace{0.2in} \nonumber\\
&=& \frac{\hat{\alpha}}{\rho}+\tau_0\vec{\nabla}s,   \mbox{  since  } \\
\vec{\nabla}s&=&\frac{1}{\mathcal{J}_I}\frac{\partial\vec{x}_I}{\partial \rho}\times\frac{\partial\vec{x}_I}{\partial \alpha}=\frac{\hat{b}_0}{1-\rho\kappa_0\cos\alpha}. \label{grad s}
\end{eqnarray}

Consequently, the tangent vectors of intrinsic coordinates can be written in terms of the gradients of the intrinsic coordinates as
\begin{eqnarray}
\frac{\partial\vec{x}_I}{\partial \rho}&=&\vec{\nabla}\rho; \\
\frac{\partial\vec{x}_I}{\partial \alpha}&=&\rho^2(\vec{\nabla}\alpha-\tau_0\vec{\nabla}s);\\
\frac{\partial\vec{x}_I}{\partial s}&=&\{(1-\rho\kappa_0\cos\alpha)^2-\rho^2\tau_0^2\}\vec{\nabla}s \nonumber\\&& -\rho^2\tau_0\vec{\nabla}\alpha.
\end{eqnarray}


\subsection{Hamiltonian for field lines \label{sec: Hamiltonian}}

This section gives the derivation of the exact Hamiltonian $\tilde{H}(\tilde{\psi},\alpha,s)=\tilde{\psi}h(\alpha,s)$ for the magnetic field lines that are in the neighborhood of an arbitrarily chosen line, called the central line, which means in the limit as the distance from the line $\rho$ goes to zero.  This Hamiltonian is determined by four functions of the distance $s$ along the central magnetic field line.

Any field that is intrinsically divergence free, such as the magnetic field, can be written \cite{Boozer:B-H} in the form 
\begin{equation}
2\pi\vec{B}= \vec{\nabla}\tilde{\psi}\times\vec{\nabla}\alpha+\vec{\nabla}s\times\vec{\nabla}\tilde{H}.  \label{B rep}
\end{equation}
This follows from writing the vector potential in the general form 
\begin{equation}
2\pi\vec{A}=\tilde{\psi}(\rho,\alpha,s)\vec{\nabla}\alpha-\tilde{H}(\rho,\alpha,s)\vec{\nabla}s+\vec{\nabla}g(\rho,\alpha,s), \label{A-int}
\end{equation} 
where $(\rho,\alpha,s)$ are the coordinates of the intrinsic coordinate system.   Any vector in three space can be described by three functions of the coordinates.  In Equation (\ref{A-int}) these are  $\tilde{\psi}(\rho,\alpha,s)$, $\tilde{H}(\rho,\alpha,s)$, and $g(\rho,\alpha,s)$.

Equation (\ref{B rep})  implies Hamiltonian equations for the magnetic field lines:
\begin{eqnarray}\frac{d\tilde{\psi}}{ds}&\equiv&\frac{\vec{B}\cdot\vec{\nabla}\tilde{\psi}}{\vec{B}\cdot\vec{\nabla}s}=- \frac{\partial \tilde{H}(\tilde{\psi},\alpha,s)}{\partial\alpha}\\ \frac{d\alpha}{ds}&\equiv&\frac{\vec{B}\cdot\vec{\nabla}\alpha}{\vec{B}\cdot\vec{\nabla}s}=\frac{\partial \tilde{H}(\tilde{\psi},\alpha,s)}{\partial\tilde{\psi}}.
\end{eqnarray}
When the functional dependence of $\tilde{H}$ is chosen to be $(\tilde{\psi},\alpha,s)$ rather than the coordinates $(\rho,\alpha,s)$ of the intrinsic coordinate system,  $\tilde{H}$ is the Hamiltonian for the magnetic field lines.  The coordinates $(\tilde{\psi},\alpha,s)$ are the canonical coordinates of  $\tilde{H}$.

The magnetic flux enclosed by a circle of radius $\rho$ is given by $\mathcal{J}_I\vec{B}\cdot\vec{\nabla}s$ integrated over $\alpha$ from zero to $2\pi$ and over $\rho$.  But, $\mathcal{J}_I\vec{B}\cdot\vec{\nabla}s=(\partial\tilde{\psi}/\partial\rho)/2\pi$, so $\oint \tilde{\psi}(\rho,\alpha)d\alpha/2\pi$ is that flux.  To the lowest non-trivial order,
\begin{equation} \tilde{\psi}=\pi B_0(s)\rho^2 \mbox{   as  } \rho\rightarrow0, \label{flux}\end{equation}
where $B_0(s)$ is the field strength along the line $\vec{x}_0(s)$.

$\tilde{H}$ to lowest non-trivial order in $\rho$ will be derived in the coordinates $(\rho,\alpha,s)$ using the expression for the current density parallel to the central magnetic field line.  This requires writing $\vec{B}$ in the covariant form
\begin{eqnarray} \vec{B}&=&B_\rho \vec{\nabla}\rho + B_\alpha \vec{\nabla}\alpha+ B_s\vec{\nabla}s \mbox{   with  }\label{cov} \\
\mu_0j_{||}&\equiv&\hat{b}_0\cdot\vec{\nabla}\times\vec{B} \nonumber\\
&=&\vec{\nabla}s\cdot(\vec{\nabla}B_\rho\times\vec{\nabla}\rho+\vec{\nabla}B_\alpha\times\vec{\nabla}\alpha) \nonumber\\
&=&\frac{1}{\mathcal{J}_I}\Big(\frac{\partial B_\alpha}{\partial \rho}- \frac{\partial B_\rho}{\partial \alpha}\Big),
\end{eqnarray}
which is be calculated in the limit $\rho\rightarrow0$.
\begin{eqnarray}
B_\alpha &=& \vec{B}\cdot\frac{\partial\vec{x}_I}{\partial\alpha}=\vec{B}\cdot\rho^2(\vec{\nabla}\alpha-\tau_0\vec{\nabla}s)  \nonumber \\
&=&\frac{\rho^2}{2\pi \mathcal{J}_I}\Big(\frac{\partial\tilde{H}}{\partial\rho} - \tau_0\frac{\partial\tilde{\psi}}{\partial \rho}\Big) \nonumber \\
&\rightarrow&\frac{\rho}{2\pi}\frac{\partial\tilde{H}}{\partial\rho}-B_0\rho^2\tau_0.
\end{eqnarray}

\begin{eqnarray}
B_\rho &=& \vec{B}\cdot\frac{\partial\vec{x}_I}{\partial\rho} =\vec{B}\cdot\vec{\nabla}\rho\nonumber \\
&=&-\frac{ \frac{\partial \tilde{H}}{\partial\alpha}+\frac{\partial \tilde{\psi}}{\partial s}}{2\pi\mathcal{J}_I}\nonumber\\
&\rightarrow&-\frac{\frac{\partial \tilde{H}}{\partial\alpha}}{2\pi\rho}
\end{eqnarray}
A derivative with respect to $s$, which means along the magnetic field, is intrinsically smaller in problems of interest compared to the derivatives with respect to $\rho$ and $\alpha$, which mean derivatives across the field.  When magnetic field lines exponentiate apart, derivatives across the field lines are exponentially larger than derivatives along the lines.

The differential equation for $\tilde{H}$  is then
\begin{eqnarray}
\nabla_\bot^2\tilde{H} &=& 4\pi B_0\left(\frac{\mu_0j_{||}}{2B_0} + \tau_0 \right), \mbox{  where   } \label{nabla^2 H}\\
\nabla_\bot^2\tilde{H} &\equiv&\frac{1}{\rho}\frac{\partial}{\partial\rho}\left(\rho\frac{\partial \tilde{H}}{\partial\rho}\right)+\frac{1}{\rho^2}\frac{\partial^2\tilde{H}}{\partial\alpha^2}
\end{eqnarray}

Writing $\tilde{H}=\pi B_0\rho^2 h(\alpha,s)$, the differential equation for $\tilde{H}$ implies
\begin{eqnarray}
&& \pi B_0\Big(4h +  \frac{\partial^2h}{\partial\alpha^2}\Big)=2\pi\mu_0j_{||}+4\pi\tau_0 B_0,  \mbox{   so } \\
&& h =  k_\omega(s)+ k_q(s)\cos(2\alpha - \varphi_q(s))  \mbox{   with  } \label{eq:h}\\
&& k_\omega \equiv  \frac{K_0(s)}{2} + \tau_0(s), \\
&&K_0 \equiv \frac{\mu_0 j_{||}}{B_0}, \hspace{0.2in} \mbox{   and  }\\
&& \tilde{H} = \tilde{\psi} h(\alpha,s). \label{near H}
\end{eqnarray}
The torsion $\tau_0(s)$ measures the departure of the central line from lying in a plane. 

The position vector of the Courant and Snyder intrinsic coordinate system, Equation (\ref{Intrinsic coordinates}), can be rewritten using the canonical coordinates of the Hamiltonian $\tilde{H}(\tilde{\psi},\alpha,s)$ of Equation (\ref{near H}),
\begin{eqnarray}
\vec{x}_{Ic}(\tilde{\psi},\alpha,s)&=&\sqrt{\frac{\tilde{\psi}}{\pi B_0(s)}} \Big(\hat{\kappa}_0(s) \cos\alpha  + \hat{\tau}_0(s) \sin\alpha \Big) \nonumber\\&&  \hspace{0.1in} +\vec{x}_0(s).
\end{eqnarray} 

In time dependent problems, it is important to have the position vector $\vec{x}_{Ic}$ given as a function of the canonical coordinates and time.  When the magnetic field is time dependent, each of the functions that appears in $\vec{x}_{Ic}(\tilde{\psi},\alpha,s,t)$ depends on time.  These functions are $B_0(s,t)$, $\hat{\kappa}_0(s,t)$, $\hat{\tau}_0(s,t)$, and $\vec{x}_0(s,t)$.


\section{Ideal evolution of neighboring lines \label{sec:ideal evolution} }

The basic field line equations were derived in Reference \cite{Boozer:B-line.sep} but not the ideal evolution equations, which will be obtained in this section. 

The evolution equations for neighboring field lines are easier to interpret if one considers a similar magnetic field representation that is not limited to vicinity of an arbitrarily chosen line.  This more general representation is given in Appendix \ref{sec:canonical coord}. 

\subsection{Electric field equations}

The evolution of a magnetic field is determined by the electric field, which can be obtained from the time-derivative of the vector potential. Equation (\ref{A-int}) for the vector potential in intrinsic coordinates implies
\begin{eqnarray}
\left(\frac{\partial \vec{A}}{\partial t}\right)_{\vec{x}}&=&-\left(\frac{\partial \tilde{H}}{\partial t}\right)_c \frac{\vec{\nabla}s}{2\pi} + \vec{u}_c\times\vec{B}+\vec{\nabla}\frac{\partial g}{\partial t}; \hspace{0.2in}\label{dA/dt}\\
\vec{u}_c&\equiv&\frac{\partial\vec{x}_{Ic}(\tilde{\psi},\alpha,s,t)}{\partial t}.  
\end{eqnarray}
The derivation of Equation (\ref{dA/dt}) is non-trivial but is given in the appendix to Reference \cite{Boozer:RMP}.  The electric field equals $-(\partial \vec{A}/\partial t)_{\vec{x}}$ minus the gradient of a potential:
\begin{equation}
\vec{E} = \left(\frac{\partial \tilde{H}}{\partial t}\right)_c \frac{\vec{\nabla}s}{2\pi}-\vec{u}_c\times\vec{B}-\vec{\nabla}(\Phi+\phi). \label{Ideal E 1}
\end{equation}

When the evolution is ideal, the electric field is also given by 
\begin{eqnarray}
\vec{E}&=& -\vec{u}_\bot\times\vec{B} -\vec{\nabla}\Phi,\label{Ideal E 2}
\end{eqnarray}
where $\vec{u}_\bot$ is the velocity of the magnetic field lines through space.  An ideal evolution of a magnetic field is uniquely described by $\vec{u}_\bot$ since $\partial\vec{B}/\partial t=\vec{\nabla}\times(\vec{u}_\bot\times\vec{B})$.

The interpretation of Equations (\ref{Ideal E 1}) and (\ref{Ideal E 2}) is clarified by a related canonical representation of the magnetic field, Appendix \ref{sec:canonical coord}, that is not limited to the vicinity of a given magnetic field line. 


\subsection{Freedom of canonical transformations}

The freedom of canonical transformations is given by $\phi$.  When Equations (\ref{Ideal E 1}) and (\ref{Ideal E 2}) are used to eliminate the electric field from the analysis:
\begin{eqnarray}
 \left(\frac{\partial \tilde{H}}{\partial t}\right)_c\frac{\vec{\nabla}s}{2\pi}&=&(\vec{u}_c-\vec{u}_\bot)\times\vec{B} +\vec{\nabla}\phi.
\end{eqnarray}


Dotting this equation with $\vec{B}$,
\begin{equation}
 \left(\frac{\partial \tilde{H}}{\partial t}\right)_c = 2\pi \frac{\partial \phi}{\partial s}. \label{dH/dt}
 \end{equation}
 
 Crossing with $\vec{\nabla}s$,
\begin{equation}
\vec{u}_c-\vec{u}_\bot = - \frac{\vec{\nabla}s\times\vec{\nabla}\phi}{B_0},  \label{u_c choice}
 \end{equation}
 
 Although $\vec{u}_\bot$ is defined by an ideal evolution, the choice of the velocity of the canonical coordinates $\vec{u}_c$ has freedom, the freedom of the canonical transformations given by $\phi$.  The choice $\vec{u}_c=\partial \vec{x}_{Ic}/\partial t$ fixes $\phi$ and through Equation (\ref{dH/dt}) the evolution of the Hamiltonian.
 
 
 \subsection{Vorticity of the field line flow}
 
 The $\hat{b}_0$ component of the vorticity of the magnetic field line flow will be shown to have an important relationship to $K_0=\mu_0j_{||}/B_0$.
 
 The vorticity of the field line flow about the central line is calculated using Equation (\ref{u_c choice}) for $\vec{u}_c-\vec{u}_\bot$: 
\begin{eqnarray}
\vec{\Omega} & \equiv & \vec{\nabla}\times \Big(\vec{u}_c(\vec{x},t)-\vec{u}_\bot(\vec{x},t) \Big) \\
&=& \frac{\vec{\nabla}s}{B_0} \nabla^2\phi - \frac{\vec{\nabla}s}{B_0}\cdot\vec{\nabla} \vec{\nabla}\phi \nonumber\\&& - \Big((\vec{\nabla}\phi) \vec{\nabla}\cdot\frac{\vec{\nabla}s}{B_0} - (\vec{\nabla}\phi\cdot\vec{\nabla})\frac{\vec{\nabla}s}{B_0}\big) \nonumber\\
&=& \frac{\hat{b}_0}{B_0} \nabla^2\phi - \frac{1}{B_0}\frac{\partial \vec{\nabla}\phi}{\partial s} \nonumber\\ && -2\frac{\vec{\nabla}\phi}{B_0^2}\frac{\partial  B_0}{\partial s} -\frac{\partial \phi}{\partial s}\left(\frac{\kappa_0\hat{\kappa}_0}{B_0} - \frac{\hat{b}_0}{B_0^2}\frac{\partial B_0}{\partial s}\right)
\end{eqnarray}
since to lowest order in $\rho$, $\vec{\nabla}s=\hat{b}_0(s)$ and $\hat{b}_0\cdot\vec{\nabla}s=1$.  The derivatives along the magnetic field, $\partial/\partial s$ are small compared to derivatives across the magnetic field---in fact exponentially smaller when the magnetic field is chaotic.  Therefore, to lowest order in $\rho$,
\begin{eqnarray}
\Omega &\equiv&\hat{b}_0\cdot\vec{\Omega} \nonumber\\
&=& \frac{\nabla_\bot^2\phi}{B_0}. \label{nabla^2 phi}
\end{eqnarray}


\subsection{Relation beween $\Omega$ and $K_0\equiv\mu_0j_{||}/B$}

Equation (\ref{nabla^2 H}) for $\nabla_\bot^2\tilde{H}$, Equation (\ref{dH/dt}) for $(\partial H/\partial t)_c$, and Equation (\ref{nabla^2 phi}) for $\Omega$ imply
\begin{eqnarray}
\frac{\partial \Omega B_0}{\partial s}&=&\frac{\partial \left(K_0 + 2\tau_0 \right)B_0}{\partial t}. \label{K ev}
\end{eqnarray}

Equation (\ref{K ev}) is a generalization of the expression $\partial\Omega/\partial s= \partial K_0/\partial t$,  Equation (4) of Reference  \cite{Boozer:j-relax} on the Alfv\'enic relaxation of variation in $j_{||}/B$ along magnetic field lines and Equation (26) of Reference \cite{Rec-example} on how simple non-turbulent footpoint motions in the solar corona can give complicated ribbons of current.  The primary difference is the inclusion of torsion $\tau_0$.  As explained in Section \ref{sec:footpoint}, including $\tau_0$ generally gives only a small correction.


\section{Reconnection due to footpoint motion \label{sec:footpoint} } 

This section extends the results on reconnection driven by the motion of the interception points of magnetic field lines with a boundary \cite{Rec-example} by using an analysis based on neighboring magnetic field lines. 

\subsection{Exponentiation of $\Delta_{max}/\Delta_{min}$ from boundary conditions}

In the solar corona, the evolution of magnetic field lines is driven by the motion of the footpoints at their two interceptions with a surface outside of the photosphere.  As with any natural flow, this flow is in general chaotic even when it is a smooth and simple function of position.  By the definition of a chaotic flow, neighboring streamlines separate exponentially in time until the separation is comparable to the longest spatial scales of the flow \cite{Rec-example}.   

When for simplicity the footpoint flow is assumed to be zero at one interception of the field lines and non-zero at the other, the boundary conditions imply the ratio of separations of neighboring magnetic field lines $\Delta_{max}/\Delta_{min}$ must be greater or equal to the separation of neighboring streamlines of the flowing footpoints.  Consequently, $\ln(\Delta_{max}/\Delta_{min}) \gtrsim t/\tau_{ev}$, where the ideal evolution time $\tau_{ev}$ is defined by the flowing footpoints.   

The existence of a spatial scale $\Delta_d$ below which magnetic field lines cannot be distinguished implies a fast magnetic reconnection over the spatial scale of the footpoint flow $a_u$ on the timescale $\sim \tau_{ev}\ln(a_u/\Delta_d)$.  What is subtle is the force-free current density in the plasma $K=\mu_0j_{||}/B$.  Boozer and Elder \cite{Rec-example} showed this current characteristically lies in thin but wide ribbons along the magnetic field lines with the maximum current density scaling as $\ln(a_u/\Delta_d)$, as predicted in Reference \cite{Boozer:rec-phys}.

What is new is that in Boozer and Elder \cite{Rec-example} the force-free current $K$ was calculated using $\partial\Omega/\partial s=\partial K/\partial t$.   The more accurate and rigorous derivation of this paper gives Equation (\ref{K ev}).  As discussed below, the difference in these two equations produces no qualitative difference in the results.


\subsection{Constraint of force balance \label{sec:force balance}}

In the solar corona, the timescale of the footpoint flow is generally  far longer than the time for shear Alfv\'en waves to propagate from one field line interception to the other.  Force balance then implies $dK_0/ds=0$.

The constraint $\vec{\nabla}\cdot\vec{j}=0$ is mathematically equivalent to 
\begin{equation}
 \vec{B}\cdot\vec{\nabla}K = \vec{B}\cdot\vec{\nabla}\times\frac{\mu_0\vec{f}_L}{B^2},
 \end{equation}
 where $\vec{f}_L \equiv\vec{j}\times\vec{B}$ is the Lorentz force.  When $dK_0/ds$ is large, the plasma inertia $\rho d\vec{v}/dt$ must balance the Lorentz force.  The $\hat{b}_0$ component of the curl of the force balance equation gives
 \begin{eqnarray}
\frac{\partial\Omega}{\partial t} = V_A^2  \frac{\partial K_0}{\partial s}, \mbox{  where   } V_A^2\equiv \frac{B_0^2}{\mu_0\rho} \label{force}
 \end{eqnarray}
 is the Alfv\'en speed \cite{Boozer:j-relax,Rec-example}.  Consequently, $\partial K_0/\partial s=0$ when the evolution of $\Omega$ is slow compared to the time for a shear Alfv\'en wave to propagate all along a magnetic field line.
 
 
 \subsection{Vorticity boundary condition}
 
 When an ideal evolution is driven by a perfectly-conducting flowing-wall, which moves with a divergence-free velocity $\vec{v}_w$ tangential to the wall, a field line strikes the wall at $\vec{x}_0(t)$, which is given by $d\vec{x}_0/dt=\vec{v}_w(\vec{x}_0,t)$.  The vorticity is 
 \begin{equation} 
 \vec{\Omega}_w\equiv\vec{\nabla}\times\Big(\vec{v}_w(\vec{x},t)-\vec{v}_w(\vec{x}_0,t)\Big),
 \end{equation} 
 where $\vec{v}_w=\hat{n}\times\vec{\nabla}f_w$, with $\hat{n}$ the unit normal to the wall.  If the wall is flat, the vorticity at the particular point is $\vec{\Omega}_w=\hat{n}(\nabla^2f_w)_{\vec{x}_0(t)}$.  The boundary condition on the vorticity about the magnetic field line that is attached to the point $\vec{x}_0(t)$ is $\Omega=(\hat{b}\cdot\hat{n}\nabla^2f_w)_{\vec{x}_0}$.
 
As discussed in Appendix \ref{sec:Spitzer ell=1}, intrinsic coordinates have a subtlety when the curvature and torsion are produced by a curl-free perturbation in which $\vec{B}/B_0 = \hat{z} + \epsilon \big(- \hat{x} \sin kz + \hat{y} \cos kz\big)$.   Even when $\epsilon\rightarrow0$, the curvature unit vector $\hat{\kappa}_0$, rotates by $2\pi$ as $z$ changes by $2\pi/k$.   This is a coordinate system oddity, not a physical effect of the relative twist of the two interceptions of the magnetic field lines, and can be removed by subtracting the $\epsilon\rightarrow0$ torsion from that given by the perturbation.  The physically relevant torsion is then $\tau_{0}\approx  - \epsilon^2/2$.
 
 
 \subsection{Slow footpoint motion}
 
 The motion of the footpoints that drive the solar corona is sufficiently slow compared to the Alfv\'en speed that $K_0\equiv \mu_0 j_{||}/B$ must be independent of $s$, but $K_0$ can depend on time. Equation (\ref{K ev}) for the evolution of $K_0$ then implies that the vorticity $\Omega$ associated with the two boundary interceptions, at $s_1$ and $s_2$, of a field line give
 \begin{eqnarray}
&& \frac{\Omega(s_2) B_0(s_2) - \Omega(s_1) B_0(s_1)}{s_2-s_1} \nonumber\\
&& \hspace{0.4in} =\frac{\partial K_0\Big<B_0\Big>}{\partial t} + 2 \frac{\partial \Big<\tau_0 B_0 \Big>}{\partial t}, \mbox{   where  }\hspace{0.4in} \\
&& \Big<B_0\Big>\equiv \frac{\int_{s_1}^{s_2} B_0 ds}{s_2-s_1}.
 \end{eqnarray}
 
 As discussed above and in Appendix \ref{sec:Spitzer ell=1}, the effective torsion $\tau_0\approx-\epsilon^2/2$, when the wobble or curvature of the central line scales as $\epsilon$, and generally gives only a small correction.  
 
 The parallel current given by $K_0$ increases by a factor of $\ln(1/\epsilon_{ni})$ before reconnection takes place.  Consequently, $\partial K_0/\partial t$ is approximately equal to $(\Omega(s_2)-\Omega(s_1))/(s_2-s_1)$ as was assumed in Reference \cite{Rec-example}.


\section{Ideal evolution of magnetic surfaces \label{sec:surface distortion} }

Reference \cite{Boozer:surf2022},  \emph{The rapid destruction of toroidal magnetic surfaces}, showed that a large ideal perturbation can contort the magnetic surfaces sufficiently that even the smallest deviation $\mathcal{E}$ from an ideal electric field produces fast magnetic reconnection.  This section uses the equations for the trajectories of field lines in the neighborhood of an arbitrarily chosen line to show that the separation of neighboring lines in the same magnetic surface and the separation of neighboring magnetic surfaces are characteristically exponential.  Reference \cite{Boozer:surf2022} did not give an explicit demonstration of the exponential behavior.

Where magnetic fields lines lie in surfaces that enclose a bounded volume, the surfaces must be toroidal with a poloidal angle $\theta$ and a toroidal angle $\varphi$.  The poloidal angle can be chosen \cite{Boozer:RMP} so 
\begin{eqnarray}
2\pi\vec{B}&=&\vec{\nabla}\psi\times\vec{\nabla}\theta+\iota(\psi) \vec{\nabla}\varphi\times\vec{\nabla}\psi 
\end{eqnarray}
The toroidal magnetic flux enclosed by a surface is $\psi$.  

An arbitrary magnetic field line, $\vec{x}_0(s)$,  can be taken to be the central line with its neighboring lines followed using the method of Section \ref{sec:near given}.  

When neighboring magnetic field lines lie in the same magnetic surface, $\psi=\psi_0$, the angle $\alpha$ of  intrinsic coordinates must obey a constraint---the separation between the lines must be orthogonal to the unit normal $\hat{n}_0=(\vec{\nabla}\psi/\big| \vec{\nabla}\psi \big|)_{\vec{x}_0}$ to the surface.   Since $\alpha$ is defined relative the curvature unit vector $\hat{\kappa}_0(s)$ of the central line, $\alpha$ must equal $\alpha_\kappa(s)$, where 
\begin{eqnarray}
&&\sin\alpha_\kappa(s)\equiv\hat{n}_0\cdot\hat{\kappa}_0.  \label{alpha-constraint}\\
&&\mbox{   Let   } \hspace{0.2in}  k_\kappa(s) \equiv \frac{d\alpha_\kappa}{ds}. \label{k_kappa}
\end{eqnarray}

The Hamiltonian of Equation (\ref{near H}) implies $d\alpha/ds = h(\alpha,s)$, so the constraint $\alpha=\alpha_\kappa(s)$ implies
\begin{eqnarray}
\frac{d\alpha_\kappa}{ds} &=& h(\alpha_\kappa,s)  \\
&=&k_\omega + k_q \cos(2\alpha_\kappa - \phi_q). \label{d alpha_kappa}
\end{eqnarray}
Using Equation (\ref{k_kappa}),
\begin{eqnarray}
&& \cos(2\alpha_\kappa - \phi_q) = \frac{k_\kappa-k_\omega}{k_q}. \label{eq:cos}
\end{eqnarray}
Since $| \cos(2\alpha_\kappa - \phi_q) |\leq1$, the $k$'s associated with a magnetic surface must satisfy $k_q^2 \geq \left(k_\kappa-k_\omega\right)^2$ when a magnetic surface exists.  A special case of this constraint was given and discussed in Section V.C of Reference \cite{Boozer:B-line.sep} for cylindrically symmetric magnetic surfaces.  In the cylindrically symmetric case, $k_{\kappa}=0$.

The separation of a neighboring line from the central line is given by
\begin{eqnarray}
\frac{d\ln\tilde{\psi}}{ds} &=& -\left(\frac{\partial h}{\partial\alpha}\right)_{\alpha=\alpha_\kappa};   \\\nonumber\\
\mbox{   Let   } \hspace{0.1in}  \tilde{\psi}&=&\tilde{\psi}_0 e^{2\sigma(s)},  \hspace{0.1in}  \mbox{     then   }\\ \nonumber\\
\frac{d\sigma}{ds} &=& -\frac{1}{2}\left(\frac{\partial h}{\partial\alpha}\right)_{\alpha=\alpha_\kappa}\\\nonumber\\
&=&  k_q \sin(2\alpha_\kappa-\phi_q) \\ \nonumber\\
&=&\pm  \sqrt{k_q^2-\left(k_\kappa-k_\omega\right)^2} \label{sigma eq}
\end{eqnarray}
with $k_q^2 \geq \left(k_\kappa-k_\omega\right)^2$.  Note that all three $k$'s are functions of the distance  $s$ along the central field-line $\vec{x}_0(s)$.  The two signs of $\sigma$ in Equation (\ref{sigma eq}) correspond to the two directions across the field lines in a given magnetic surface.  Along a single field line, $d\sigma/ds$ can be positive for some and negative for other  $s$.

The separation between neighboring lines in the same magnetic surface is $\Delta_{||}=\rho$ at $\alpha=\alpha_\kappa$ with $\tilde{\psi}=\pi B_0 \rho^2$, so
\begin{eqnarray}
\Delta_{||}(s) = \Delta_{||}(0) \sqrt{\frac{B_0(0)}{B_0(s)}} e^{\sigma(s)}.
\end{eqnarray}

The flux in a tube $\tilde{\psi}$ is proportional to $B_0(s)\Delta_{||}\Delta_{\bot}$, where $\Delta_{\bot}(s)$ is the separation between surfaces.  Since $\oint (d\ln\tilde{\psi}/ds)d\alpha=0$, the product $B_0(s)\Delta_{||}\Delta_{\bot}$ is independent of $s$ as was found in Reference \cite{Boozer:surf2022};
\begin{eqnarray}
\Delta_\bot(s) = \Delta_\bot(0) \sqrt{\frac{B_0(0)}{B_0(s)}} e^{-\sigma(s)}.
\end{eqnarray}
Neighboring magnetic surfaces are close to each other where neighboring lines in the same surface are far apart and vice versa.

Although the natural separation of neighboring field lines is exponential, the extent of the exponentiation depends on the situation.  In axisymmetric equilibria, only pairs of field lines that are started arbitrarily close to an X-point can undergo an arbitrarily large number of exponentiations in separation.  The case of cylindrically symmetric magnetic surfaces, Section V.C of Reference \cite{Boozer:B-line.sep}, illustrates how exponentiation is avoided by $k_q^2 = \left(k_\kappa-k_\omega\right)^2$.  It is also explained how X-points and a separatrix arise when this constraint is weakly broken.

These results on magnetic surfaces have two obvious applications.  The first is the determination of the numerical limitations of codes that calculate the breakup of magnetic surfaces.  The sudden loss, $\sim 1~$ms, of a large fraction of the magnetic surfaces is a common feature of tokamak disruptions.   Non-ideal effects on the magnetic evolution are extremely small in the central regions of ITER or future power plants.  As discussed in \cite{Boozer:surf2022,Huang:2022}, obtaining an adequate resolution for a study of ideal perturbations is challenging even in the highly simplified case of steady state perturbations that preserve helical symmetry.  A method of resolution analysis based on following individual magnetic field lines in a numerical simulation is discussed in \cite{Boozer:surf2022}.  The second application is the extrapolation of results that can be obtained numerically to the situations in which they are needed.  It appears unlikely that sufficient numerical resolution can be obtained to realistically follow magnetic-surface breakup in large tokamaks.  It will, therefore, be necessary to extrapolate from the situations that can be can be numerically studied to those for which information is needed.  An understanding of how the breakup actually occurs is required if reliable extrapolations are to be made. 

In a tokamak, the current density $j_I$ required to sustain an island can be shown to be approximately equal to the spatially averaged axisymmetric current density $\bar{j}_{eq}$ enclosed by the rational surface on which the island is centered.   In principle, higher current densities can be obtained by suddenly turning on a strong perturbation.

The ideal response to an ideal magnetic perturbation was discussed in Appendix B of \cite{Boozer:surf2022}.  The ratio of $j_\xi$, the maximum current density that arises to preserve the  constraints of the ideal evolution and prevent the opening of an island,  to $\bar{j}_{eq}$ can be estimated using results from that Appendix:
\begin{equation}
\frac{j_\xi}{\bar{j}_{eq}}\approx \frac{\xi_d}{r_s} \frac{t}{\tau_A},
\end{equation}
where $\xi_d$ is the ideal displacement of magnetic surfaces well away from the rational surface, which has a radius $r_s$.  The timescale $\tau_A$ is proportional to the time it takes a shear Alfv\'en wave to propagate toroidally around the plasma.  In the infinite time limit, a delta-function current must be located on the rational surface to preserve the ideal-evolution constraints \cite{Huang:2022}.

Unfortunately, numerical results are not available for the $j_\xi/\bar{j}_{eq}$ ratio versus time, even in the relatively simple case of helical symmetry.  Such calculations are needed in the full toroidal case to understand how tearing modes and the breakup of magnetic surfaces proceed in a realistically evolving tokamak plasma.  Traditional reconnection theory would predict that resistivity causing surface breakup in $\sim 1~$ms in plasmas in which the time scale for resistive dissipation of the current is tens of minutes,  as in ITER, would require $j_\xi/\bar{j}_{eq}\sim10^5$, which not only seems implausible but too much time would be required to achieve the required current density.  A current density that is large compared to the equilibrium current density, which is typically 1~MA/m$^2$ in tokamaks, would have a spectacular effect on electron runaway to relativistic energies.


\section{Discussion \label{sec:discussion} }

The emphasis of magnetic reconnection studies should shift from the use of traditional theory to the use of the intrinsic properties of Faraday's law---not only because of its greater range of applications but also because of its sound mathematical foundations:  (1) Fast magnetic reconnection is intrinsic to Faraday's law. (2) The parallel electric field required by the dominant reconnection mechanism is many orders of magnitude smaller than that required by traditional reconnection theory. (3)  A large parallel electric field is not necessary for particle acceleration. (4) Even simple flows that are non-turbulent produce extremely complicated current sheets and particle acceleration patterns. (5) Turbulence slows the speed of reconnection in comparison to smooth flows that are on the scale of the reconnection region.  

These five points have been made in publications that are reviewed in this paper.  This paper focuses on the intrinsic properties that can be better understood by examining the behavior of magnetic field lines in the neighborhood of an arbitrarily chosen line---called the central line.  

In addition to the development of a new formalism, two specific results are new.  (1) Equation (\ref{K ev}) gives the precise relation between the time derivative of the current density along a magnetic field line, $K_0\equiv\mu_0j_{||}/B$, and the derivative along the line of the flow vorticity, $\Omega$, about the line.  This relation is fundamental to the Alfv\'enic flattening of $j_{||}/B$ along a magnetic field line \cite{Boozer:j-relax} and to the production of complicated current sheets in the plasma volume by simple boundary motions \cite{Rec-example}. Unlike earlier results, it includes the effects of the torsion of the central magnetic field line and of the variation in field strength along the line.  (2)  Equation (\ref{sigma eq}) gives the exponential variation in the separation of neighboring magnetic field lines that lie in the same magnetic surface as well as the exponential variation in the separation of neighboring magnetic surfaces.  Although earlier work \cite{Boozer:surf2022} showed that an ideal evolution could make the variation in separation between neighboring surfaces arbitrarily large, it had not explicitly shown this variation is exponential. 

Section \ref{sec:surface distortion} derives the maximum current density that can arise in response to a magnetic perturbation.  Although this current density goes to infinity in the infinite-time limit, its increase is too slow to explain the fast breakup of magnetic surfaces in tokamak disruptions without an exponential  enhancement of the ratio of separations of neighboring field lines, $\Delta_{max}/\Delta_{min}$.

Appendix \ref{sec:acceleration} gives an important result for collisionless charged particles.  In an evolving magnetic field, the maximum kinetic energy in the direction of the magnetic field generically increases exponentially in time. 

 
\section*{Acknowledgements}
This work was supported by the U.S. Department of Energy, Office of Science, Office of Fusion Energy Sciences under Award Numbers DE-FG02-95ER54333, DE-FG02-03ER54696, DE-SC0018424, and DE-SC0019479.

\section*{Data availability statement}

Data sharing is not applicable to this article as no new data were created or analyzed in this study.



\appendix


\section{Exponential particle acceleration \label{sec:acceleration}}

When a magnetic field is evolving---even evolving ideally---the maximum parallel kinetic energy of collisionless charged particles naturally increases exponentially with time.  As will be shown, this follows from Hamiltonians of the $H(p,q,t)$ form generically having chaotic trajectories, which means neighboring pairs of $(p,q)$ trajectories separate exponentially in time.  Hamiltonians in the actual six dimensional phase space $H(\vec{p},\vec{q},t)$ of charged particle motion share the generic property of being chaotic.

The exponential increase in energy has been known, but the derivations are complicated \cite{Drake:2017,Boozer:acc}.  Chaos-induced acceleration is known \cite{Drake:2016} to often be more efficient than acceleration by the non-ideal parallel electric field $\mathcal{E}\vec{\nabla}\ell$ even in traditional reconnection theories.

For simplicity, assume the gyroradius of the accelerated particle is so small that the drifts across the magnetic field lines are negligible.  The velocity of the particle enters only through $v_{||}$, its velocity component along $\vec{B}$, and $\mu\approx mv_\bot^2/2B$, its invariant magnetic moment.  The Hamiltonian for the particle is 
\begin{equation}
H_p(p_{||},\ell,t)=  \frac{p_{||}^2}{2m} + \mu B(\ell,t),  \label{particle H}
\end{equation} 
where the momentum $p_{||}=mv_{||}$ and $\ell$ is the distance along the magnetic field line.  When the magnetic field strength is independent of time, the energy of the particle $H_p$ is a constant of the motion.

The energy of the particle is not conserved when $H_p$ has an explicit dependence on time.  Using Hamilton's equations, $dp_{||}/dt=-\partial H_p/\partial\ell$ and $d\ell/dt=\partial H_p/\partial p_{||}$, one finds the rate of change of the particle's energy $dH_p/dt=\partial H_p/\partial t$.  

The magnetic field strength $B(\ell,t)$ is generally time dependent even when the magnetic field evolves ideally, which means $\mathcal{E}=0$.  The implication is that particles undergo acceleration and deceleration in their parallel velocity even when $\mathcal{E}=0$.

Pairs of neighboring trajectories of a generic Hamiltonian $H(p,q,t)$ separate from each other exponentially in time.  Consequently, when the field strength in Equation (\ref{particle H}) depends on time, the maximum particle energy should increase exponentially in time for collisionless particle---even when the magnetic evolution is ideal, $\mathcal{E}=0$.

Although the chaos that is characteristic of Hamiltonians of the $H(p,q,t)$ form is an essential concept in  the theory of fast magnetic reconnection and particle acceleration, it does not appear to be broadly appreciated within the plasma community.  It can be illustrated even in the Hamiltonian for a harmonic oscillator, $H=p^2/2m + kq^2/2$, when the spring constant $k$ depends weakly on time.  When $k=k_0$, a constant, a trajectory has the form $p=p_0 \cos(\omega t)$, $q = (\omega/k_0)p_0 \sin(\omega t)$, $ \omega^2\equiv k_0/m$, and the energy is $H=p_0^2/2m$ and constant.  When $k=k_0\big(1-\epsilon \sin(2\omega t)\big)$, then as $\epsilon\rightarrow0$, the rate of change of $H$ averaged of a period is $\Big<dH/dt\Big>=(\epsilon\omega/2)H$.  A far more involved but realistic example is the simple pendulum with a length $\ell_p$ that is a complicated function of time.

The extent of the exponentiation in the parallel energy is highly dependent on the situation: (1) How long does the particle stay in the region in which the variation in $B(\ell,t)$ is large?  (2) Does the $\ell$ variation of $B$ have scales as short as $v_{||}/\omega_c$, which would cause diffusion in the magnetic moment $\mu$?  (3) Do kinetic instabilities arise that cause an increase in $\mu$?  Detailed, situation specific, calculations are required to obtain information on the resulting particle-energy distributions.


\section{Magnetic canonical coordinates \label{sec:canonical coord} }

This appendix discusses a general canonical form for a magnetic field that is closely related to the representation using intrinsic coordinates.

\subsection{A general canonical form for $\vec{B}$}

The explicit representation of a magnetic field in terms of a Hamiltonian and its canonical coordinates was first published \cite{Boozer:B-H} in 1983.  A variant on the 1983 form will be useful for understanding the evolution of magnetic fields.

Except at locations at which $\vec{B}=0$, the trajectories of magnetic field lines at a fixed time $t$ can be represented by a Hamiltonian $H(\psi,\alpha,\ell,t)$, where $\ell$ is the distance along a line and $\alpha$ an angle such that $\vec{B}\times\vec{\nabla}\alpha\neq0$.
\begin{eqnarray}
&& 2\pi \vec{B} = \vec{\nabla}\psi\times \vec{\nabla}\alpha +\vec{\nabla}\ell \times\vec{\nabla}H(\psi,\alpha,\ell,t). \label{B-rep}\\
&& \frac{d\psi}{d\ell} \equiv \frac{\vec{B}\cdot\vec{\nabla}\psi}{\vec{B}\cdot\vec{\nabla}\ell}=-\frac{\partial H(\psi,\alpha,\ell,t) }{\partial\alpha } \label{H_1}\\
&& \frac{d\alpha}{d\ell} \equiv \frac{\vec{B}\cdot\vec{\nabla}\alpha}{\vec{B}\cdot\vec{\nabla}\ell}=\frac{\partial H(\psi,\alpha,\ell,t) }{\partial\psi}; \label{H_2}
\end{eqnarray}
$(\psi,\alpha,\ell)$ are the canonical coordinates of the Hamiltonian $H$.  Knowledge of the spatial locations of field lines also requires knowledge of the position vector $\vec{x}(\psi,\alpha,\ell,t)$ of the canonical coordinates
\begin{equation}
\vec{x} =x(\psi,\alpha,\ell,t)\hat{x} + y(\psi,\alpha,\ell,t)\hat{y} + z(\psi,\alpha,\ell,t)\hat{z}.
\end{equation}

Equation (\ref{B-rep}) for $\vec{B}$ can be easily derived by first writing the vector potential in the general form
\begin{equation}
2\pi \vec{A}= \psi \vec{\nabla}\alpha + H \vec{\nabla}\ell +\vec{\nabla}g, \label{A-rep}
\end{equation}
where the gauge $g$ can be used represent any dependence of $\vec{A}$ on the third coordinate.  The curl gives Equation (\ref{B-rep}).  One can show that $\psi$ is the magnetic flux enclosed by a constant-$\psi$ curve at fixed $\ell$.


\subsection{Ideally evolving magnetic field}

Obtaining the time derivative of the vector potential of Equation (\ref{A-rep}) at a fixed position $\vec{x}$ is non-trivial to determine, but this derivative is given in the Appendix to Reference \cite{Boozer:RMP},
\begin{eqnarray}
\left(\frac{\partial \vec{A}}{\partial t}\right)_{\vec{x}}&=&-\left(\frac{\partial H}{\partial t}\right)_c \frac{\vec{\nabla}\ell}{2\pi} + \vec{u}_c\times\vec{B}+\vec{\nabla}\tilde{s}; \hspace{0.2in} \\
\vec{u}_c&\equiv&\left(\frac{\partial \vec{x}}{\partial t}\right)_c,
\end{eqnarray}
where the subscript ``$c$" means the canonical coordinates $(\psi,\alpha,\ell)$ are held constant, $\vec{u}_c$ is the velocity of these coordinates through space, and the function $\tilde{s}(\vec{x},t)$ is arbitrary due to the freedom of gauge. 

The electric field differs from $-(\partial \vec{A}/\partial t)_{\vec{x}}$ by the gradient of a single-valued potential.  Consequently, when the evolution is ideal, $\mathcal{E}=0$, the general representation for the electric field is
\begin{eqnarray}
\vec{E}&=&- \vec{u}_\bot\times\vec{B} -\vec{\nabla}\Phi  \label{E1}\\
&=& \left(\frac{\partial H}{\partial t}\right)_c\frac{\vec{\nabla}\ell}{2\pi} -\vec{u}_c\times\vec{B}-\vec{\nabla}(\Phi+\phi). \hspace{0.2in} \label{E2}
\end{eqnarray}

The potential $\phi$ represents the freedom of canonical transformations.  Subtracting Equation (\ref{E1}) from Equation (\ref{E2}), one finds that
\begin{eqnarray}
\left(\frac{\partial H}{\partial t}\right)_c\frac{\vec{\nabla}\ell}{2\pi} = (\vec{u}_c-\vec{u}_\bot)\times\vec{B} + \vec{\nabla}\phi. \label{gen H ev}
\end{eqnarray}
The component of the equation parallel to $\vec{B}$ is
\begin{eqnarray}
\left(\frac{\partial H}{\partial t}\right)_c&=& 2\pi \left(\frac{\partial \phi}{\partial \ell}\right)_c,
\end{eqnarray}
where $(\partial\phi/\partial\ell)_c\equiv(\vec{B}/B)\cdot \vec{\nabla}\phi$.  The components perpendicular to $\vec{\nabla}\ell$ are obtained by noting 
\begin{eqnarray}
\vec{\nabla}\ell\times\vec{\nabla}\phi &=& - \vec{\nabla}\ell\times\{(\vec{u}_c-\vec{u}_\bot)\times\vec{B} \} \\
&=&-(\vec{u}_c-\vec{u}_\bot)B +\vec{B}( \vec{u}_c-\vec{u}_\bot)\cdot\vec{\nabla}\ell \hspace{0.2in}\nonumber\\
\end{eqnarray}
since $\vec{B}\cdot\vec{\nabla}\ell=B$. The velocity of the canonical coordinates can be taken to be related to the magnetic field line velocity by
\begin{equation}
\vec{u}_c= \vec{u}_\bot -\frac{\vec{\nabla}\ell\times\vec{\nabla}\phi}{B}.  \label{u_c eq}
\end{equation}



\section{Spitzer's stellarator fields \label{sec:Spitzer}}

Spitzer \cite{Spitzer:1958} gave forms for magnetic fields in a cylinder that illustrate the production of a rotational transform by either torsion, which is his $\ell=1$ case, or by a quadrupole field, which is his $\ell=2$, case.  In both cases $\vec{B} = B_0\hat{b}$ with $\hat{b} = \hat{z} + \hat{z} \times \vec{\nabla}\mathcal{H}$.  The magnetic field produced by $\mathcal{H}$ was assumed to be a small perturbation, so to lowest order the distance along a magnetic field line is $z$.


\subsection{Spitzer's form for $\ell=1$ \label{sec:Spitzer ell=1} }

For the $\ell=1$ case, Spitzer chose $B_\theta=A \cos(\theta-kz)$ in cylindrical coordinates with $A$ a constant. This is equivalent to $\mathcal{H}=\epsilon r\cos(\theta-kz)$ since $B_\theta=\partial \mathcal{H}/\partial r$.  The analysis is simpler in Cartesian coordinates in which
\begin{eqnarray}
\mathcal{H}&=&\epsilon  \big(x \cos kz + y \sin kz\big) ; \\
\vec{b} &=& \hat{z} + \epsilon \big(- \hat{x} \sin kz + \hat{y} \cos kz\big);\\
\vec{\kappa} &=& \frac{\epsilon k}{\sqrt{1+\epsilon^2}} \hat{\kappa};\\
\hat{\kappa}&=&-\hat{x}\cos kz - \hat{y} \sin kz; \\
\frac{d\hat{\kappa}}{dz} &=& k \big( \hat{x} \sin kz -\hat{y} \cos kz\big) =-(\kappa\hat{b}+\tau\hat{\tau});\\
\vec{\tau}&=& \frac{k}{\sqrt{1+\epsilon^2}} \hat{\tau};\\
\hat{\tau}&=&\frac{\hat{x} \sin kz -\hat{y} \cos kz -\epsilon\hat{z}}{\sqrt{1+\epsilon^2}}.
\end{eqnarray}

Remarkably, the torsion does not vanish when $\epsilon=0$.  This can be intepreted by considering the intrinsic coordinates of Section \ref{sec:intrinsic coord}.  In intrinsic coordinates, the angle $\alpha$ is defined relative to the direction $\hat{\kappa}$, and during each period of length $L\equiv2\pi/k$ the angle $\alpha$ advances by $\tau L=2\pi$ in the limit as $\epsilon\rightarrow0$.  The effective torsion, or twist of the magnetic field lines, is $\tau_{eff}=\tau(\epsilon) - \tau(0) \approx  - \epsilon^2/2$.


\subsubsection{Spitzer's form for $\ell=2$}

Spitzer \cite{Spitzer:1958} gave $B_\theta=A r \cos(2\theta-kz)$ for an $\ell=2$ magnetic field, which is equivalent to 
\begin{eqnarray}
\mathcal{H}&=&\frac{1}{2}\epsilon r^2\cos(2\theta-kz)\nonumber\\
&=&\frac{\epsilon r^2}{2}\Big\{(\cos(2\theta)\cos kz + \sin(2\theta)\sin kz\Big\}\nonumber\\
&=&\frac{\epsilon}{2} \Big\{(x^2-y^2)\cos kz + 2xy\sin kz\Big\}.
\end{eqnarray}
This is a rotating quadrupole magnetic field, the same field as the one that appears in Equation (\ref{eq:h}).

The twist of the field lines over a period given by this perturbation is $\epsilon^2$.  The perturbation produces no curvature or torsion of the central field line, the line at $r=0$.




\cleardoublepage


\begin{thebibliography}{99}

\bibitem{Boozer:rec-phys} A. H. Boozer, \emph{Magnetic reconnection and thermal equilibration}, Phys. Plasmas \textbf{28}, 032102 (2021).

\bibitem{Aref:1984} H. Aref, \emph{Stirring by chaotic advection}, Journal of Fluid Mechanics \textbf{143}, 1 (1984).

\bibitem{Boozer:surf2022} A. H. Boozer, \emph{The rapid destruction of toroidal magnetic surfaces}, Phys. Plasmas \textbf{29}, 022301  (2022). 

\bibitem{Boozer:coordinates} A. H. Boozer, \emph{Plasma equilibrium with rational magnetic surfaces}, Phys. Fluids \textbf{24}, 1999 (1981).

\bibitem{X-null} A. H. Boozer, \emph{Magnetic reconnection with null and X-points}, Phys. Plasmas \textbf{26}, 122902 (2019). 

\bibitem{Elder:2021} T. Elder and A. H. Boozer, \emph{Magnetic nulls in interacting dipolar fields},
J. Plasma Phys. \textbf{87}, 905870225 (2021). 

 \bibitem{Newcomb} W. A. Newcomb, \emph{Motion of magnetic lines of force}, Ann. Phys. \textbf{3}, 347 (1958).

\bibitem{Parker-Krook:1956} E. N. Parker and M. Krook, \emph{Diffusion and severing of magnetic lines of force}, Ap. J. \textbf{124}, 214 (1956).

\bibitem{Rec-example} A. H. Boozer and T. Elder, \emph{Example of exponentially enhanced magnetic reconnection driven by a spatially bounded and laminar ideal flow}, Phys. Plasmas \textbf{28}, 062303 (2021).

\bibitem{Boozer:B-line.sep} A. H. Boozer, \emph{Separation of magnetic field lines}, Phys. Plasmas \textbf{19}, 112901 (2012).

\bibitem{Schindler:1988} K. Schindler, M. Hesse, and J. Birn, \emph{General magnetic reconnection, parallel electric-fields, and helicity}, Journal of Geophysical Research---Space Physics \textbf{93}, 5547 (1988).

\bibitem{Hesse-Cassak2020} M. Hesse and P. A. Cassak, \emph{Magnetic reconnection in the space sciences: Past, present, and future}, Journal of Geophysical Research: Space Physics, \textbf{125}, e2018JA025935 (2020). 

\bibitem{plasmoid} D. A. Uzdensky, N. F. Loureiro, and A. A. Schekochihin, \emph{Fast Magnetic Reconnection in the Plasmoid-Dominated Regime}, Phys. Rev. Lett. \textbf{105}, 235002 (2010).




\bibitem{Kuhn} T. S. Kuhn, \emph{The Structure of Scientific Revolutions} (University of Chicago Press, Chicago and London, 1962).

\bibitem{Britannica:Kuhn} Britannica, The Editors of Encyclopaedia (2021, July 14). ``Thomas S. Kuhn." Encyclopedia Britannica. https://www.britannica.com/biography/Thomas-S-Kuhn.

\bibitem{Drake:2016} J. T. Dahlin, J. F. Drake, and M. Swisdak, \emph{Parallel electric fields are inefficient drivers of energetic electrons in magnetic reconnection}, Phys. Plasmas \textbf{23}, 120704 (2016).  

\bibitem{Drake:2017} J. T. Dahlin, J. F. Drake, and M. Swisdak, \emph{The role of three-dimensional transport in driving enhanced electron acceleration during magnetic reconnection}, Phys. Plasmas \textbf{24}, 092110 (2017).

\bibitem{Boozer:acc} A. H. Boozer, \emph{Particle acceleration and fast magnetic reconnection}, Phys. Plasmas  \textbf{26}, 082112 (2019).

\bibitem{Lazarian:1999} A. Lazarian and E. T. Vishniac, \emph{Reconnection in a weakly stochastic field}, Ap. J. \textbf{517}, 700 (1999).

\bibitem{Eyink:2011} G. L. Eyink, A. Lazarian, E. T. Vishniac, \emph{Fast magnetic reconnection and spontaneous stochasticity}, Ap. J. \textbf{743}, 51 (2011).

\bibitem{Eyink:2015} G. L. Eyink, \emph{Turbulent general magnetic reconnection}, Ap. J \textbf{807} 137 (2015).
 
\bibitem{Matthaeus:2015} W. H. Matthaeus, M. Wan, S. Servidio, A. Greco, K. T. Osman, S. Oughton, and P. Dmitruk, \emph{Intermittency, nonlinear dynamics and dissipation in the solar wind and astrophysical plasmas}, Phil. Trans. R. Soc. A \textbf{373} 20140154 (2015).

\bibitem{Matthaeus:2020} S. Adhikari, M. A. Shay, T. N. Parashar, P. S. Pyakurel, W. H. Matthaeus, D. Godzieba, J. E. Stawarz, J. P. Eastwood, and J. T. Dahlin,  \emph{Reconnection from a turbulence perspective}, Phys. Plasmas \textbf{27}, 042305  (2020).

\bibitem{Lazarian:2020rev} A. Lazarian, G. L. Eyink, A. Jafari, G. Kowal, H. Li, S-Y Xu, and E. T. Vishniac, \emph{3D turbulent reconnection: Theory, tests, and astrophysical implications}, Phys. Plasmas \textbf{27}, 012305 (2020).

\bibitem{Priest:2016} E. Priest, \emph{MHD structures in three-dimensional reconnection}, volume \textbf{427}, page 101, \emph{Book Series Astrophysics and Space Science Library},  \emph{Magnetic reconnection: concepts and applications}, (Springer International Publishing 2016, edited by WalterGonzalez and Eugene Parker)

\bibitem{Reid:2020} J. Reid, C. E. Parnell, A. W. Hood, and P. K. Browning, \emph{Determining whether the squashing factor, Q, would be a good indicator of reconnection in a resistive MHD experiment devoid of null points}, Astronomy and Astrophysics \textbf{633}, A92 (2020).

\bibitem{Courant:1958}  E. D. Courant and H. S. Snyder, \emph{Theory of the alternating-gradient synchrotron}, Ann. Phys. \textbf{3}, 1 (1958). See Appendix B.

\bibitem{Boozer:RMP} A. H. Boozer, \emph{Physics of magnetically confined plasmas}, Rev. Mod. Phys. \textbf{76}, 1071 (2004).

\bibitem{Huang:2022}  Y.-M. Huang, S. R. Hudson, J. Loizu, Y. Zhou, and A. Bhattacharjee, \emph{Numerical study of $\delta$-function current sheets arising from resonant magnetic perturbations}, Phys. Plasmas \textbf{29}, 032513 (2022). 

\bibitem{Boozer:B-H} A. H. Boozer, \emph{Evaluation of the structure of ergodic fields}, Phys. Fluids \textbf{26}, 1288 (1983).

\bibitem{Boozer:j-relax} A. H. Boozer, \emph{Flattening of the tokamak current profile by a fast magnetic reconnection with implications for the solar corona}, Phys. Plasmas \textbf{27}, 102305 (2020).

\bibitem{Spitzer:1958} Lyman Spitzer, \emph{The stellarator concept}, Phys. Fluids \textbf{1}, 253 (1958).




 \end{thebibliography}
\end{document}